\documentclass[journal=jacsat,manuscript=article]{achemso}
\usepackage{graphicx}
\usepackage{dcolumn}
\usepackage{multirow}
\usepackage{bm}

\usepackage[utf8]{inputenc}
\usepackage[T1]{fontenc}
\usepackage{mathptmx}
\usepackage{etoolbox}
\usepackage{physics}
\usepackage{color} 
\usepackage{hyperref}
\usepackage{amsmath}
\usepackage{booktabs}       
\usepackage{threeparttable}
\usepackage{appendix}

\usepackage[inline]{enumitem}

\usepackage[version=3]{mhchem} 

\newcommand{\im}{\textup{i}}

\author{Yong-Tao Ma}
\affiliation{%
Department of Chemistry, School of Science and Research Center for Industries of the Future, Westlake University, Hangzhou, Zhejiang 310024, China
}%
\altaffiliation{Contributed equally to this work}

\author{Rui-Hao Bi}%
\affiliation{%
Department of Chemistry, School of Science and Research Center for Industries of the Future, Westlake University, Hangzhou, Zhejiang 310024, China
}%
\altaffiliation{Contributed equally to this work}

\author{Wenjie Dou}
\affiliation{%
Department of Chemistry, School of Science and Research Center for Industries of the Future, Westlake University, Hangzhou, Zhejiang 310024, China
}%
\alsoaffiliation{%
Institute of Natural Sciences, Westlake Institute for Advanced Study, Hangzhou, Zhejiang 310024, China
}%
\alsoaffiliation{%
Key Laboratory for Quantum Materials of Zhejiang Province, Department of Physics, School of Science and Research Center for Industries of the Future,
Westlake University, Hangzhou, Zhejiang 310024, China
}%
\email{douwenjie@westlake.edu.cn}

\date{\today}

\title[An \textsf{achemso} demo]{Orbital Surface Hopping from Orbital 
Quantum-Classical Liouville Equation for 
Nonadiabatic Dynamics of Many-electron Systems}

\abbreviations{IR,NMR,UV}
\keywords{American Chemical Society, \LaTeX}

\SectionNumbersOn

\begin{document}

\begin{tocentry}

\begin{center}
\includegraphics[height=4.5cm]{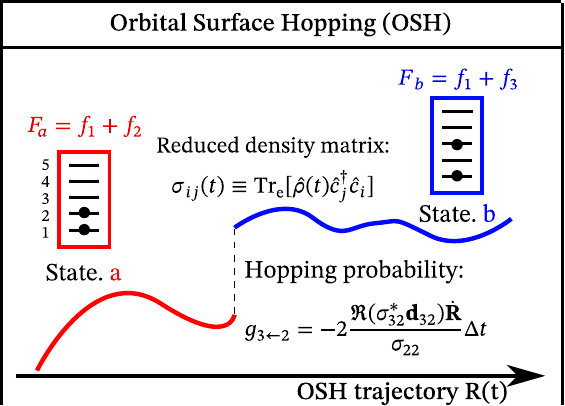}
\end{center}

\end{tocentry}

\begin{abstract}

Accurate simulation the many-electronic nonadiabatic dynamics process at metal surfaces
remains as a significant task. In this work,
we present an orbital surface hopping (OSH) algorithm rigorously derived 
from the orbital quantum classical Liouville equation (o-QCLE) to deal with 
nonadiabatic dynamics for many-electron systems. 
This OSH algorithm closely connects with the popular Independent Electron
Surface Hopping (IESH) method, which has shown remarkable success
in addressing these nonadiabatic phenomena, except that electrons hop between orbitals.
We compare OSH with IESH approach and benchmark these two algorithms
against the surface hopping method with a full Configuration Interaction (FCI) wavefunction. 
Our approach shows strong agreement with IESH and FCI-SH results for molecular orbital populations and
kinetic energy relaxation and in high efficiency, demonstrating the ability of the new OSH method in capturing key aspects of many-electronic nonadiabatic dynamics.
\end{abstract}
\maketitle
\section{Introduction}

Nonadiabatic effects on metal surfaces, such as vibrational energy relaxation and electron transfer, play a crucial 
role in many physical and chemical processes, such as inelastic 
collision\cite{wodtke2004,Wodtke04b,Wodtke20a,Wodtke20b,tully2000rev}, chemisorption \cite{HAdsporption-Bunermann2015, 
AgSurfNODiss_Krger2016, AgSurfHClDiss-Geweke2020}, electrochemistry \cite{PCET-Lam2020, Surface-Lee2022}, heterogeneous 
catalysis \cite{NiMethane-Luo2016, H-Dorenkamp2018}, and molecular junctions. \cite{Mjunction-Ke2021, CISS-Teh2022} 
Accurately describing these effects, particularly the interaction between nuclear motion and electronic degrees of freedom, 
presents a significant challenge due to the severe breakdown of the Born-Oppenheimer approximation. Moreover, the metal electronic 
orbitals form a continuum, facilitating surface electron-hole pair excitations. Consequently, treating these systems fully quantum 
mechanically would require considering an astronomical number of excited states.These factors make full quantum mechanic methods like Multi 
Configuration Time Dependent Hartree (MCTDH)\cite{Beck20001}, Hierarchical Quantum Master Equation (HQME)\cite{Qiang2019} 
impractical for such systems.

For efficient simulations of more realistic metal interface systems, a mixed quantum-classical approach is necessary. Indeed, 
a variety of methods such as electron friction theory (EF),
\cite{joseph2017,james1993,Suhl1975,Newns1995,HGM1995,Joseph2015,Joseph2016,Felix2011,Daniel2012,PERSSON1980} 
independent electron surface hopping (IESH) method\cite{tully2009j,tully2012fd}, Mapping Mode\cite{Nandini2024}, 
broached classical master equation (BCME) methods\cite{QCMEalg-D2015,QCMEderive-D2015} have recently been used to study 
the dynamics of metal surface. Among these methods, the EF and IESH methods are the most popular and widely used. 
Although EF theory is straightforward to implement and has seen significant development in recent years,
\cite{Joseph2018, Wang2021, Burghardt2022prl,Burghardt2022pra, Jorg2018,Bin2020jpcc,Bin2022jpcl,Reinhard2023, StephenJ2023, 
dou2023jcp,dou2023prb,dou2024jpcc} it is inherently limited to the weak coupling regime. Specifically, its applicability
requires low nuclear momenta and weak nuclei-electron interactions. Furthermore, when analyzing vibrational energy 
dissipation, EF theory can accurately predict the average vibrational energy but fails to provide the correct
vibrational state distribution.\cite{tully2012fd}

To address the limitations of the EF method, Tully's group 
developed the IESH method in 2009.\cite{tully2009j} 
The original IESH method was applied to the 
inelastic scattering of NO-Au(111) system, where it successfully reproduces trapping probabilities and equilibrium vibrational 
energy distributions that align well with experimental results.\cite{tully2009s} This demonstrates that IESH can capture key experimental 
trends with reasonable accuracy. Furthermore, it is computationally efficient due to the independent electron assumption, 
making it a highly practical and effective model. Since its development, IESH has been widely applied to open-shell species 
on metal surfaces \cite{tully2009jpcc}.Despite its successes, IESH remains an ans\"{a}tz without rigorous proof.
In this article, we demonstrate why the independent electron ans\"{a}tz of IESH works by directly analyzing the many-body quantum dynamics of a non-interacting system. The main theoretical result is an orbital quantum classical Liouville equation (o-QCLE), which can naturally lead to an orbital surface hopping (OSH) algorithm. In addition, we perform full configuration space surface hopping dynamics for small systems to demonstrate both OSH and IESH capture the essential many body dynamics despite being evolved with single-particle equations of motions.

This article is organized as follows. Section~\ref{sec:Theroy} comprises a comprehensive description of methods and theory used in this paper, where Sec.~\ref{subsection:o-QCLE} presents the derivation of orbital quantum classical Liouville equation. The o-QCLE can naturally leads to the OSH approach, which presents in Sec.~\ref{sebsection:OSH}; In Sec.~\ref{subsection:IESH} we reconstructed the IESH method based on o-QCLE; Sec.~\ref{subsection:FCI} outlines the algorithm of Full Configuration Interaction Surface Hopping (FCI-SH) algorithms. Section~\ref{sec:results} presents the numerical comparison of IESH, OSH, and FCI-SH over two systems, representing thermal equilibrium (Sec.~\ref{subsection:results_eq}) and high vibrational state (Sec.~\ref{subsection:results_non_eq}), respectively. 
Finally, we summarize this work in Section~\ref{sec:conclusion}.

\section{Theory and Methodology}\label{sec:Theroy}

\subsection{The Orbital Quantum Classical Liouville Equation}\label{subsection:o-QCLE} 
Consider a general many-electron Hamiltonian describing the coupled electron-nuclear motion, given by
\begin{equation}
    \hat{H}_\text{total} = \hat{T}_\text{n} + \hat{H}_\text{el}(\bm{r},\bm{R}),
\end{equation}
where $\hat{T}_\text{n} \equiv \hat{\bm{P}} \bm{M}^{-1} \hat{\bm{P}} / 2$ represents the nuclei kinetic operator, 
$\hat{H}_\text{el}$ represents the electronic Hamiltonian that depends on electronic ($\bm{r}$) and 
nuclear ($\bm{R}$) coordinates. Suppose the electrons are noninteracting, such that $\hat{H}_\text{el}$ can be described 
by quadratic terms  
\begin{equation}\label{eqn:many_body_Hmol}
    \hat{H}_\text{el}(\bm{r},\bm{R}) = \sum_{ij}^m h_{ij}(\bm{R}) \hat{d}_i^{\dagger} \hat{d}_j + U(\bm{R}),
\end{equation}
with $\hat{d}_i^{\dagger}$ ($\hat{d}_j$) denotes the creation (annihilation) operator for the $i$ ($j$)-th orbital, 
$h_{ij}(\bm{R})$ is the single-bady Hamiltonian
and $U(\bm{R})$ represents electronic state independent potential energy. This system comprises $m$-spin orbitals, 
which are occupied by $n$ electrons distributed among these orbitals.

The many-body quantum dynamics of this molecular system is governed by the Liouville-von Neumann (LvN) equation, that
\begin{equation}\label{eqn:lvn}
    \frac{\partial \hat{\rho}}{\partial t} = -\frac{\im}{\hbar} \comm{\hat{H}_\text{total}}{\hat{\rho}(t)},
\end{equation}
where $\hat{\rho}(t)$ denotes the total many-body density operator. To proceed, we first define the single-body density operator $\hat{\sigma}$ whose matrix elements are given by $\hat{\sigma}_{kl} \equiv \Tr_\text{e}(\hat{\rho} d^{\dagger}_l d_k)$. We can show that the single-body density operator satisfies (See detailed derivations in Appendix.~\ref{app:lvn})
\begin{equation}\label{eqn:single_body_lvn}
    \frac{\partial \hat{\sigma}}{\partial t} = -\frac{\im}{\hbar} \comm{\hat{H}^{orb}_\text{total}}{\hat{\sigma}(t)},
\end{equation}
Here $\hat{H}_{total}^{orb}$ represents the total Hamiltonian in the single-body basis set $\{\ket{i}, i=1, \dots, m\}$:
\begin{equation}\label{eqn:single_body_Hmol}
    \hat{H}^{orb}_\text{total} = \hat{T}_\text{n} +U (\hat{\bm{R}}) + \sum_{ij}^m h_{ij}(\bm{R}) \ketbra{i}{j},  
\end{equation}
again, $U(\bm{R})$ represents the pure nuclear potential energy.
The single-body LvN Eq.~\ref{eqn:single_body_lvn} indicates that for a noninteracting many electronic system, 
the dynamics are \emph{exactly} encoded in the single-body Hamiltonian, as long as the system can be described 
by quadratic terms (Eq.~\ref{eqn:many_body_Hmol}). 

We proceed by deriving a mixed quantum-classical Liouville equation for the orbital density matrix. To do so, we can apply the partial Wigner transformation, which transfers the nuclear degree of freedom from Hilbert space to phase space. For an arbitrary operator $\hat{O}(t)$, Wigner transformation reads: 
\begin{equation}
    \hat{O}_{W}(R,P,t) =\int dY e^{\frac{-iP\cdot Y}{\hbar}}\bra{R-\frac{Y}{2}}\hat{O}(t)\ket{R+\frac{Y}{2}},
\end{equation} 

and for single-body density matrix $\hat{\sigma}$,
\begin{equation}
    \hat{\sigma}_{W}(R,P,t) = (2 \pi \hbar)^{-3N} \int \dd \bm{Y} e^{\frac{-i\bm{Y}\cdot \bm{P}}{\hbar}}
    \mel{\bm{R} - \frac{\bm{Y}}{2}}{\hat{\sigma}(t)}{\bm{R} + \frac{\bm{Y}}{2}}, 
\end{equation}
where $3N$ is the size of the nuclear degrees of freedom. The trace of $\hat{\sigma}_{W}(R,P,t)$ over electronic degrees of freedom equals the number of electrons, $m$, as this is a many-electron system. By applying the partial Wigner transform rules to Eq.~\ref{eqn:single_body_lvn} and truncating the Wigner-Moyal operator to the first order \cite{QCLE-Kapral1999}, we obtain the orbital quantum-classical Liouville equation (o-QCLE)

\begin{equation}\label{eqn:qcle_general}
\begin{split}
    \frac{\partial \hat{\sigma}_{W}(\bm{R},\bm{P},t)}{\partial t} = &-\frac{i}{\hbar}[\hat{h}_{W},\hat{\sigma}_{W}] -\sum_{a} \frac{P_{a}}{M_{a}}\frac{\partial\hat{\sigma}_{W}}{\partial R_{a}} \\ 
  &+\frac{1}{2}\sum_{a}\big(\{\hat{h}_{W},\hat{\sigma}_{W}\}-\{\hat{\sigma}_{W},\hat{h}_{W}\}\big) + \sum_{a}\frac{\partial U}{\partial R_{a}}\frac{\partial\hat{\sigma}_{W}}{\partial P_{a}}.    
\end{split}  
\end{equation}
Here the Poisson bracket is 
\begin{equation}
  \{\hat{A}(\bm{R,P}),\hat{B}(\bm{R,P})\} = \frac{\partial \hat{A}}{\partial \bm{R}} \cdot \frac{\partial \hat{B}}{\partial \bm{P}} -  \frac{\partial \hat{A}}{\partial \bm{P}} \cdot \frac{\partial \hat{B}}{\partial \bm{R}}.
\end{equation}

Finally, o-QCLE can be projected onto the adiabatic basis. We first diagonalize the single-body Hamiltonian to get the adiabatic 
orbital energy $\epsilon_j(\bm{R})$ and corresponding basis $\ket{\phi_{j}(\bm{R})}$:
\begin{equation}
   h(\bm{R}) \ket{\phi_{j}(\bm{R})} = \epsilon_{j}(\bm{R}) \ket{\phi_{j}(\bm{R})} . 
\end{equation}

Sandwiching both sides of Eq.~\ref{eqn:qcle_general} by $\mel{\phi_j(\bm{R})}{\cdot}{\phi_k(\bm{R})}$, we obtain

\begin{equation}\label{eqn:qcle_adiabatic}
\begin{split}
     \frac{\partial \hat{\sigma}^{jk}_{W}(\bm{R},\bm{P},t)}{\partial t} = &-\im \omega^{jk}\sigma^{jk}_{W}(\bm{R},\bm{P},t) + \sum_{l} \sum_{a}\frac{P_{a}}{M_{a}}(\sigma^{jl}_{W}d_{lk}^{a}-d_{jl}^{a}\sigma^{lk}_{W}) \\  
    &- \sum_{a}\frac{P_{a}}{M_{a}} \pdv{\sigma_{W}^{jk}}{R_a} -\frac{1}{2}\sum_{l}\sum_{a} \left( F_{a}^{jl} \pdv{\sigma_{W}^{lk}}{R_{a}} + \pdv{\sigma_{W}^{jl}}{P_{a}} F_{a}^{lk} \right)  + \sum_{a} \pdv{U}{R_a} \pdv{\sigma^{jk}_{W}}{P_a},      
\end{split}
\end{equation}
where $\omega^{jk}=(\epsilon_{j}-\epsilon_{k})/\hbar$. We have used the fact that:
\begin{equation}
 \bra{\phi_{j}(\bm{R})}\frac{\partial \hat{\sigma}_{W}(t)}{\partial\bm{R}}\ket{\phi_{k}(\bm{R})} = \frac{\partial \sigma^{jk}_{W}(t)}{\partial \bm{R}}
 -\sum_{l}(\sigma^{jl}_{W}\bm{d}_{jl}-\bm{d}_{lk}\sigma^{lk}_{W}),
\end{equation}
here the nonadiabatic coupling matrix elements are given as: 
\begin{equation}
    d_{jk}^{a}=\bra{\phi_{j}(\bm{R})}\frac{\partial}{\partial R_a}\ket{\phi_{k}(\bm{R})} = -\frac{F^{jk}_{a}}{\hbar\omega^{jk}},
\end{equation}

and the force can be calculated under Hellman Feynman equation  
\begin{equation}
F^{jk}_{a} =  \bra{\phi_{j}(\bm{R})}[\partial h/\partial R_{a}] \ket{\phi_{k}(\bm{R})}.
\end{equation}

\subsection{Orbital surface hopping method (OSH)}\label{sebsection:OSH}

With o-QCLE (Eq.~\ref{eqn:qcle_adiabatic}), we propose a orbital surface hopping (OSH) algorithm. 
QCLE is the starting point of many mixed quantum-classical algorithms,
\cite{aug-Ehrenfest-Subotnik2010,derive-Subotnik2013, MASH-Mannouch2023} and its connection with surface 
hopping has been thoroughly discussed in Refs.~\cite{MJSH-Nielsen2000,derive-Subotnik2013,shperspective-Kapral2016} 
Similarly, o-QCLE (Eq.~\ref{eqn:qcle_adiabatic}) also implies a surface hopping algorithm which is called OSH.
Without considering decoherence,\cite{derive-Subotnik2013} the OSH algorithm can be outlined step-by-step as follows.

\begin{enumerate}
    \item We start by sampling the initial wavepacket with a swarm of trajectories. The initial position ($\bm{R}$), 
    momentum ($\bm{P}$) and occupied oribtals are sampled according to the problem being investigated.  
    In each trajectory, nuclear position ($\bm{R}$), 
    nuclear momenta ($\bm{P}$), and adiabatic single-particle density matrix ($\sigma$) are propagated.  In addition, an index array
    $\vec{\lambda} = \{\lambda_1, \dots, \lambda_n\}$ is used to track the occupation of $n$ electrons in $m$ adiabatic orbitals. 
    Specifically, $\sigma$ can be initialized by either occupying diabatic or adiabatic orbitals:
    \begin{itemize}
        \item Diabatic occupation. Initially, suppose $n$ diabatic orbitals, $\vec{\xi} = \{\xi_1, \dots, \xi_n\}$, are occupied. 
        Then, the initial diabatic density matrix is given by 
        \[\sigma^\text{diabatic}_{ij} =\left\{ 
        \begin{array}{ll}
        \delta_{ij}  & i \in \vec{\xi} \\
        0            & i \notin \vec{\xi} 
        \end{array}
        \right. .\]
        Adiabatic density $\sigma$ can be obtained by $\sigma = U^{\dagger} \sigma^\text{diabatic} U$. The adiabatic 
        index array $\vec{\lambda}$ is then stochastically sampled from adiabatic populations $\text{diag}(\sigma)$.
        \item Adiabatic occupation. For given adiabatic occupation configuration $\vec{\lambda}$, $\sigma$ is given by
        \[\sigma_{ij} =\left\{ 
        \begin{array}{ll}
        \delta_{ij}  & i \in \vec{\lambda} \\
        0            & i \notin \vec{\lambda} 
        \end{array}
        \right. .\]
    \end{itemize}
    Despite the initialization scheme, the initial density represent a mixed state whose trace is $n$.
    \item \label{algo:two} Between time $t$ and $t + \Delta t$, propagate $\bm{R}$, $\bm{P}$ by the following equations of motion:
    \begin{gather}
        \dot{\bm{R}} = \frac{\bm{P}}{M}, \\
        \dot{\bm{P}} = -\pdv{U}{\bm{R}} + \sum_{i=1}^{n} \bm{F}_{\lambda_i, \lambda_i}.   
    \end{gather}
    Instead of being propagated on potential energy surface of the \emph{active state}, the nuclei experience the single-particle forces of the occupied orbitals.
    $\sigma$ is propagated with a smaller time step $\Delta t'$under the equation of motion for single-particle density matrix,
    \begin{equation}
       \dot{\sigma}_{ij}(t) = -\im \omega^{jk} \sigma_{ij}(t) - \frac{\bm{P}}{M} \cdot \sum_k (\bm{d}_{ik} \sigma_{kj} - \sigma_{ik}\bm{d}_{kj}).\label{Eq:eomo} 
    \end{equation}

    \item At every smaller time step $\Delta t'$, we evaluate the hopping probabilities from each occupied orbital to the unoccupied orbitals. 
    Using the Fewest Switches Surface Hopping (FSSH) \cite{tully1990} scheme, the hopping probability from occupied orbital $\lambda_i$ to orbital $j$ is,
    \begin{equation}
        g_{j \gets \lambda_i} = \left\{
        \begin{array}{ll}
            \max \left\{-2\Re(\sigma^{*}_{j \lambda_i}\dot{\bm{R}}\cdot\bm{d}_{j \lambda_i} ) \Delta t'/ \sigma_{\lambda_i \lambda_i}, \ 0 \right\} & j \notin \vec{\lambda} \\
            0 & j \in \vec{\lambda}
        \end{array}
        \right..
    \end{equation}
    The stochastic hopping algorithm is identical to that of FSSH, \cite{tully1990} albeit hopping will be considered for each occupied electron. 
    To minimize hopping, no more than one electron is allowed to transition to a different orbital per nuclear propagating time step.
    \item When a hop occurs between $\lambda_i$ and $j$, the total energy must be conserved. \cite{derive-Subotnik2013,shperspective-Kapral2016} 
    Momenta are rescaled in the direction of the single-particle derivative coupling $\bm{d}_{j\lambda_i}$. \cite{schiffer94} 
    If the nuclei do not have enough kinetic energy, the hop is frustrated and this hopping attempt is ignored.

    \item Return to step~\ref{algo:two}.
\end{enumerate}

    Similar to the original FSSH algorithm, physical quantities that are exclusively nuclear operators (e.g. position, 
    momentum, kinetic energy and etc.) or electronic operators diagonal in the adiabatic representation (e.g. adiabatic orbital populations) 
    can be directly computed via averaging over trajectory. Diabatic orbital populations, however, is evaluated with the density matrix approach:\cite{diabpop-Landry2013,amber2022}
    \begin{equation}\label{eq:cmd}
        P_a = \sum_{i=1}^{n} \left(\abs{U_{a \lambda_i}}^2 + \sum_{j=1}^{m} \sum_{k=j}^{m} 2\Re(U_{aj} \sigma_{jk} U_{ak}^*) \right). 
    \end{equation}

Overall, we have presented an efficient surface hopping algorithm for non-interacting multiple electrons system. 
This OSH algorithm completely considers orbitals and occupations rather than electronic state, hence will be very efficient for large systems. 
Readers familiar with the Independent Electron Surface Hopping (IESH) will notice some similarity between OSH and IESH. 
We will compare these two methods in Sec.~\ref{subsection:IESH}.

\subsection{IESH method}\label{subsection:IESH}

We can reconstruct the IESH method starting from o-QCLE. As derived in Sec.~\ref{subsection:o-QCLE}, 
the dynamics are exactly encoded in the single-body Liouville equation, as long as the electronic 
interaction can be described by quadratic terms. Each electron in this system can evolve independently 
under the time-dependent Schodinger equation or single body density matrix as Eq.~\ref{Eq:eomo}, 
which is almost the same as our OSH approach. 

The main difference between IESH and OSH is that electrons hop between states. The hopping rate between different states is achieved by constructing
the many electron states with occupied orbitals.\cite{tully2009j}  If $b^{+}_{j}$ and $b_{j}$ are the creation and annihilation operators for 
a particle in adiabatic orbital basis $\ket{\phi_{j}(R)}$. Then the many-electron eigenstate can be obtained by
populating $N_{e}$ of these one-electron orbitals in one slater determinant.

\begin{equation}
    \ket{\bm{j}} = b^{+}_{j_{Ne}},\dots,b^{+}_{j_{2}}b^{+}_{j_{1}}\ket{0} = |\phi_{j_{1}}\phi_{j_{2}},\dots,\phi_{j_{Ne}}|.
\end{equation}

 Analogy the definition of FSSH\cite{tully1990}, hopping probability from state $\ket{\bm{j}}$ to state $\ket{\bm{l}}$ can be written as:

\begin{equation}
    g_{\bm{jl}} = \text{max}\left\{\frac{\Delta t B_{\bm{lj}}}{A_{\bm{jj}}},0\right\}, 
\end{equation}
in which,
\begin{equation}
    B_{\bm{jl}} = \text{-2Re} (A^{*}_{\bm{jl}}\dot{R}\cdot\bra{j_{i}}\bm{d(R)}\ket{l_{i}}). 
\end{equation}
The $j_{i}$ and $l_{i}$ are the electron occupied orbitals in states $\bm{j}$ 
and $\bm{l}$ respectively. The subscript $i$ for the $i^{th}$ electron. Here we follow Tully's definition \cite{tully2009j} 
using this structure to remove the antisymmetry of electrons.  
It is easy to verify (See detailed derivations in Appendix.~\ref{app:nac}) that the nonadiabatic coupling is 
zero unless only one pair of electron orbitals different from each other, which means only single electron hops need considered in hopping rate calculation.

The density matrix elements $A^{*}_{\bm{jl}}$ and $A_{\bm{jj}}$ are calculated with "general overlap method"\cite{Poshusta-overlap-1991}, 
\begin{equation}
  A_{\bm{jl}} = \braket{\bm{j}}{\Phi}\braket{\Phi}{\bm{l}}.   
\end{equation}
Here the total wavefunction $\ket{\Phi}$ is a single Slater determinant state,
\begin{equation}
    \ket{\Phi} = |\psi_{1}\psi_{2},\dots,\psi_{Ne}|.
\end{equation}
Then the many electron inner product $\braket{\bm{j}}{\Phi}$ is given by the determinant value of the overlap matrix of occupied orbitals:
\begin{equation}
   \braket{\bm{j}}{\Phi} = |S|,\ \  S_{ij} = \braket{\phi_{j_{i}}}{\psi_{j}}. 
\end{equation}

Generally, the computational cost of these determinants is proportional to $N^{2}$. $N$ is the dimension of the determinant. 
These will be the bottleneck of IESH approach if we calculate them directly.\cite{metal-Gardner2023} Detailed discussion is given in the following section.

\subsection{Full Configuration Interaction Surface Hopping} \label{subsection:FCI}
As discussed in Section \ref{subsection:IESH}, only one Slater determinant is used for the total many-electron wave function
in the IESH method. To include the correlation energy of different electron occupation, we borrow the idea
from electronic structure theory, the exact total many-electron wavefunction is defined as a linear combination of $N_{e}$ 
electron trial function:\cite{szabo-1996}
\begin{equation}
    \ket{\Phi}= c_{0}\ket{\Psi_{0}} + \sum_{a r}c_{a}^{r}\ket{\Psi_{a}^{r}} + \sum_{\substack{a<b\\ r<s}}c_{ab}^{rs}\ket{\Psi_{ab}^{rs}} + 
    \cdot\cdot\cdot, 
\end{equation}
where $\ket{\Psi_{0}}$ is the Hartree Fock (HF) like Slate determinant with lowest energy orbitals occupied.
$\ket{\Psi_{a}^{r}}$ are the Slate determinants involving single excitation, $\ket{\Psi_{ab}^{rs}}$ are the State determinants involving double excitation.

In this FCI-SH method, a sort of electrons evolve under many-body sch$\ddot{o}$dinger equation.
\begin{equation}
     \im\hbar\ket{\dot{\Psi}_{i}}=H_{el}(\bm{R(t)})\ket{\Psi_{i}}.
\end{equation}

The evolution of density matrix also determined by the quantum Liouville equation (Eq.3).
Under partial Wigner transformation and the adiabatic representation $\ket{\Phi_{J}(\bm{R})}$, 
the final quantum-classical Liouville equation of FCI density is written as   

\begin{equation}
\begin{split}
     \frac{\partial \hat{\rho}^{jk}_{W}(\bm{R},\bm{P},t) }{\partial t} = &-\frac{i}{\hbar}(E_{j}-E_{k})\hat{\rho}^{jk}_{W} + \sum_{l} \frac{P}{M}(\rho^{jl}_{w}D_{lk}-D_{jl}\rho^{lk}_{w}) \\ 
     &- \frac{P}{M}\frac{\partial \rho_{w}^{jk}}{\partial R}-\frac{1}{2}\sum_{l}\big(F_{w}^{jl}\frac{\partial \rho_{w}^{lk}}{\partial P} + \frac{\partial \rho_{w}^{jl}}{\partial P}F_{w}^{lk} \big). 
\end{split}
\end{equation}
Start from this QCLE, we can implies the FCI-SH algorithm step-by-step as in Sec.~\ref{sebsection:OSH}.
It worth noting that the hop probability from surface $\bm{j}$ to surface $\bm{l}$,  $g_{\bm{jl}}$ is also defined as
\begin{equation}
    g_{\bm{jl}} = \text{max}\left\{\frac{\Delta t B_{\bm{lj}}}{A_{\bm{jj}}},0\right\}, 
\end{equation}
and,
\begin{equation}
\begin{split}
    B_{\bm{jl}} &= \text{-2}\text{Re} (A^{*}_{\bm{jl}}\dot{R}\cdot\bra{\Psi_{j}}\bm{d(R)}\ket{\Psi_{l}}) \\
                &= \text{-2}\text{Re} (A^{*}_{\bm{jl}}\dot{R}\cdot\bra{j_{i}}\bm{d(R)}\ket{l_{i}}). 
\end{split}
\end{equation}
Which is exactly the same as the hopping probability equation in IESH approach. This indicates that 
the IESH approach has considered all electron excitation. The comparison of the dynamics results between IESH and FCI-SH would show clearly 
the effect of many-body wavefunction to the dynamic process, 
especially, when this wavefunction is coupled with a nuclear trajectory and 
ultrafast electron transfer happens in the nuclei movement.

Finally, as mentioned in \ref{sebsection:OSH},
to include decoherence and to use information of unoccupied state, the population of diabetic state is calculated using 
the correct density matrix (CMD) method\cite{Brian2013,amber2022} as Eq.~\ref{eq:cmd}. For the evolution processes of the IESH and FCI-SH are similar as the OSH approach. 
The details have shown in Sec~.\ref{sebsection:OSH}, so we will not repeat here.

\section{Results and discussions}\label{sec:results}
In this work, we focus on two questions: \begin{enumerate*}
    \item Has any many-body effects been missed in the non-adiabatic dynamic simulation when using a single-particle approaches such as OSH and IESH.
    \item Can the new simpler OSH approach achieve comparable, or even better results than IESH? 
\end{enumerate*}
To address these questions, we study two nonelastic collision process between a molecule and metal surface. 
One is that the molecule is under thermal equilibrium before collision, which is a simplified model used to
study the electron transfer rate in the chemisorption. The other is that the molecule is highly vibration excited, 
which can show the state-to-state energy transfer mechanism in the fast electron transfer process.         

For both systems, we employ the Newns-Anderson model, which has been successfully applied to study non-adiabatic effects at gas-metal interfaces.\cite{tully2009j,tully2009s} Numerous reference results using IESH are also available.\cite{ouyang16,miao18,amber2022} For smaller systems, we use Full Configuration Space Surface Hopping approach as a many-body benchmark. For larger systems, comparisons are limited to OSH and IESH.

The Newns-Anderson Hamiltonian with one molecular orbital and a continuum of metal electron orbitals is given as:

\begin{equation}\label{eqn:newns_anderson}
\begin{split}
    \hat{H}_\text{el}(\bm{R}) =  &U_{0}(\bm{R}) + (U_{1}(\bm{R})-U_{0}(\bm{R}))c^{\dagger}_{a}c_{a}+\int_{E_{-}}^{E_{+}} d\epsilon (\epsilon c^{\dagger}_{\epsilon}c_{\epsilon}) \\ 
    &+\int_{E_{-}}^{E_{+}} d\epsilon (V(\epsilon;\bm{R}) c^{\dagger}_{a}c_{\epsilon}+V(\epsilon;\bm{R})^{*} c^{\dagger}_{\epsilon}c_{a}),
\end{split}    
\end{equation}
where $U_{0}(\bm{R}), U_{1}(\bm{R})$ are the potential energy surfaces of the neutral and anionic molecule, respectively.  
$V(\epsilon;\bm{R})$ are the couplings between molecular and metal orbitals. 
$c^{\dagger}_{\epsilon} (c_{\epsilon})$ and $c^{\dagger}_{a} (c_{a})$ are the fermionic creation (annihilation)
operators of metal and molecule orbital, respectively.

To explicitly simulate this model, we discretize the integrals in Eq.~\ref{eqn:newns_anderson}. Specifically, 
the wide-band spectral density function is used and the hybridization function,
\begin{equation}   
    \Gamma(\epsilon; \bm{R}) \equiv 2 \pi \sum_k \abs{V(\epsilon, \bm{R})}^2 \delta(\epsilon - \epsilon_k),
\end{equation}
is assumed to be a constant $\Gamma$ across the interval $E_{-} = -W / 2 + \mu$ and $E_{+} = W / 2 + \mu$. Here, $W$ and $\mu$ are the bandwidth and the chemical potential, respectively. The trapezoid discretization procedure is adopted following Ref.~\cite{miao18}, as it is reported to be numerically more stable\cite{miao18} and equally accurate as Gauss-Legendre procedure for small bandwidths.\cite{metal-Gardner2023} For $n_\text{b}$ metal orbitals, the discretized energies ($\epsilon_k$) and the couplings ($V_{ak}$ ) are given as
\begin{equation}    
\begin{gathered}
     \epsilon_k = \mu - \frac{W}{2} + W \frac{k-1}{n_\text{b} - 1}, \\
      V_{ak} = \sqrt{\frac{\Gamma}{2\pi\rho}},
\end{gathered}
\end{equation}
where $k=1, \dots, n_\text{b}$ and the one electron state density $\rho = n_\text{b} / W$. 

For the potential energy surfaces, we consider a 1D double well potential following Refs.~\cite{ouyang16,miao18,amber2022,metal-Gardner2023}

\begin{equation}
    U_0(R) = \frac{1}{2} M \omega^{2} R^{2}, 
\end{equation}
and linear orbital energy, 
\begin{equation}
   U_1(R) - U_0(R) = -M \omega^2 g R + \frac{1}{2} M \omega^2 g^2 + \Delta G.
\end{equation}
Here, the displacement parameter $g$ is related to reorganization energy by $E_\text{r} = \frac{1}{2} M \omega^2 g^2.$


Throughout Sec.~\ref{sec:results}, the occupation of the metal orbitals are prepared at "zero temperature", 
i.e., all the diabatic orbitals below Fermi level $\mu$ are occupied. The impurity molecular orbital, on the other hand, is unoccupied.
Sec.~\ref{subsection:results_eq} shows the thermal equilibrium results, hence the trajectories are initialized at thermal equilibrium in the $U_0$ well. 
In this case, an additional nuclear friction $\gamma_\text{ext} = 2 \omega$ is applied to all trajectories to include the phononic energy dissipation on the surface.\cite{miao18} 
Sec.~\ref{subsection:results_non_eq} shows the vibrational relaxation results, where the trajectories are initialized at a high vibrational state
($\nu = 16$) by the Wigner distribution sampling \cite{wigner_function_Sun2010} of the $U_0$ harmonic potential. 
Finally, parameters used in Sec.~\ref{subsection:results_eq} and Sec.~\ref{subsection:results_non_eq} are summarized in Table~\ref{tab:model_parameters}.

\begin{table}[htbp]
    \caption{\label{tab:model_parameters}Parameters used in the two systems}
    \centering
    \begin{threeparttable}
    \begin{tabular}{c c c}
    \toprule
        \multirow{2}{*}{Parameters\tnote{a}} & \multicolumn{2}{c}{Systems} \\
           & \multicolumn{1}{c}{Equilibrium\tnote{b}} & \multicolumn{1}{c}{Relaxation\tnote{c}}  \\
        \midrule
           $M$                        & $2.0 \times 10^{3}$   & 29164.4   \\
        $\omega$                      & $2.0 \times 10^{-4}$  & $9.0 \times 10^{-3}$        \\
        $\gamma_\text{ext}$           & $4.0 \times 10^{-4}$  &  0 \\
        $\Gamma$                      & $1.0 \times 10^{-4}$  & 0.03675 \\
             g                        & 20.6097               & 0.1   \\
        $\Delta G$                    & $-3.8 \times 10^{-3}$ & 0.09356 \\
        $k_{B}$T                      & $9.5 \times 10^{-4}$  & ---   \\
        Time step (fs)                & 0.25                  & 0.005                \\
        $N_\text{traj}$               & 500                   & 5000                 \\
    \bottomrule
    \end{tabular}
    \begin{tablenotes}
    \item[a] All physical parameters are expressed in atomic units unless otherwise specified.
    \item[b] Adopted from Ref.~\cite{miao18}
    \item[c] Approximated from Ref.\cite{pes-Meng2022} See SI for details.
    \end{tablenotes}
    \end{threeparttable}
\end{table}

\subsection{Equilibrium system simulation}\label{subsection:results_eq}
The 1D Newns-Anderson model (Eq.~\ref{eqn:newns_anderson}) is widely used to model equilibrium charge transfer between adsorbed molecules 
and metal surfaces.\cite{ouyang16} Among the methods compared in previous studies,\cite{ouyang16,miao18,amber2022,metal-Gardner2023} IESH stands 
out for its accuracy and adaptability to realistic systems. However, its computational cost becomes prohibitive when many bath orbitals are 
included.\cite{miao18,metal-Gardner2023} In this subsection, we show that our OSH algorithm achieves the same results as IESH but with significantly improved efficiency.  

\begin{figure}[htbp]
    \centering
    \includegraphics[width=0.45\linewidth]{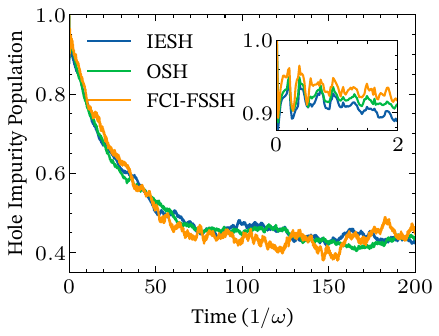}
    \caption{Impurity hole population dynamics as a function of time for OSH, IESH and FCI-SH. 
    The inset shows the short-time behavior of the same dynamics. $n_\text{b} = 10$. $N_\text{traj}=128$ for FCI-SH.}
    \label{fig:pop_eq_methods_compare}
\end{figure}

Figure~\ref{fig:pop_eq_methods_compare} illustrates the impurity hole population ($1 - \expval{c_a^{\dagger} c_a}$) dynamics, 
where both OSH and IESH agree well with FCI-SH. We highlight that despite being \emph{completely} single-particle, 
OSH shows good agreement with FCI-SH in short-time dynamics, equilibrium population, and the overall population relaxation rate. 
Our result here numerically validates both OSH and the IESH ans\"{a}tz, at least for a smaller system ($n_\text{b}=10$). 
Applying FCI-SH to larger systems becomes impractical because the configuration space scales combinatorially 
with $n_\text{b}$ (462 for $n_\text{b}=10$). It is worth noting that a similar comparison between FCI-SH and IESH was conducted by 
Pradhan and Jain~\cite{amber2022}, but their results differ from ours, showing considerable disagreement.

\begin{figure}[htbp]
    \centering
    \includegraphics[width=0.8\linewidth]{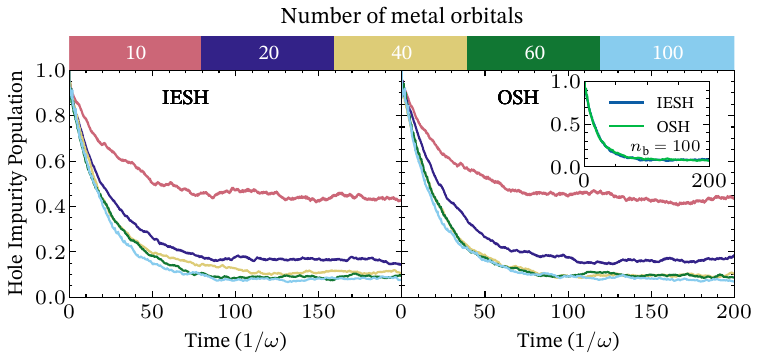}
    \caption{Convergence of the population dynamics over $n_\text{b}$. The left and right panels show results for IESH and OSH, 
    respectively. Line colors correspond to the number of bath metal orbitals, as indicated in the top color palette. 
    The inset in the right panel compares IESH and OSH when $n_\text{b} = 100$.}
    \label{fig:pop_eq}
\end{figure}

Figure~\ref{fig:pop_eq} shows the convergence of hole population dynamics as we increase orbitals $n_\text{b}$. Notice that OSH agrees very well with IESH, 
and both results are consistent with FIG. 2 (a) of Ref.~\cite{miao18}. Here, OSH converges slightly faster with respect to $n_\text{b}$. 
Notably, the simulation cost of OSH is much lower than that of IESH especially for large systems, where we observe a 4 time speed up 
when $n_\text{b} = 100$. The remarkable speedup of OSH arises from its simplicity: it only requires orbital density matrix elements $\sigma_{ij}$ 
to calculate the hopping rate,
whereas IESH must evaluate the determinant of overlap matrices for each possible hop. This well-known limitation of IESH which necessitates 
optimized numerical algorithm for hopping probability \cite{miao18} or upper bound filtering \cite{metal-Gardner2023} to mitigate, 
but OSH inherently avoids this issue. A comparison of the efficiency between OSH and IESH is provided in the Supporting Information (SI).

\subsection{Vibrational relaxation simulation }\label{subsection:results_non_eq}

In this subsection, we showcase the performance of OSH in modeling the relaxation of a highly excited vibrational mode near a metal surface. 
Specifically, we investigate the relaxation of \ce{NO} ($\nu = 16$) on a \ce{Au}(111) surface, a system previously studied using IESH by 
various groups.\cite{tully2009j,metal-Gardner2023,bin24} For simplicity, our simulations focus solely on the \ce{NO} stretching mode. 
The double-well potential used in this study is parametrized based on Ref.~\cite{pes-Meng2022} (see Supporting Information for details).

\begin{figure}[htbp]
    \centering
    \includegraphics[width=0.8\linewidth]{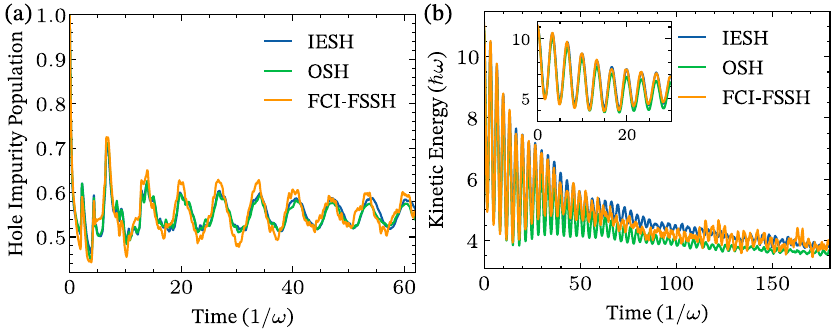}
    \caption{
    Comparison of OSH, IESH, and FCI-FSSH in modeling the relaxation of \ce{NO} ($\nu = 16$) at \ce{Au} (111) 
    surface. (a) Impurity hole population dynamics over time. (b) Kinetic energy evolution over time. 
    $n_\text{b} = 10$. $N_\text{traj}=512$ for FCI-FSSH.
    }
    \label{fig:fci_compare}
\end{figure}

Similarly as in Sect.~\ref{subsection:results_eq}, we start by comparing OSH and IESH with FCI-SH for a smaller 
$n_\text{b}$. Figure~\ref{fig:fci_compare} (a) shows consistent agreement among these methods, validating 
OSH can predict the correct electron transfer dynamics even when the nuclear system is highly excited. 
Figure~\ref{fig:fci_compare} (b) depicts the kinetic energy relaxation dynamics of the $\ce{NO}$ molecule. 
In particular, all three methods predict identical short ($<20/\omega$) and long ($>12/\omega$) time behavior 
of the kinetic energy. However, OSH predicts faster kinetic energy relaxation in the middle portion and relaxes 
to a slightly lower final kinetic energy than both IESH and FCI-SH.

\begin{figure}[htbp]
    \centering
    \includegraphics[width=0.8\linewidth]{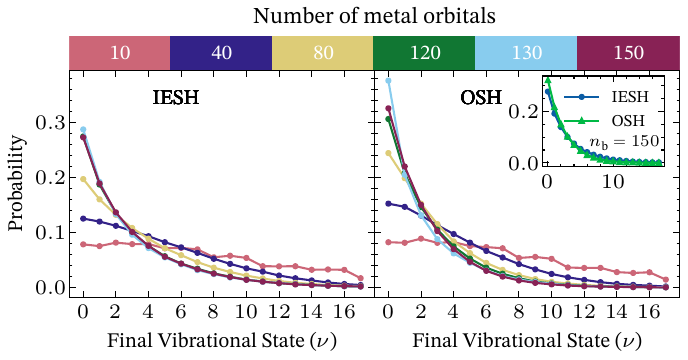}
    \caption{Convergence of final vibrational state $\nu$ over $n_\text{b}$. The left and right panels 
    show results for IESH and OSH, respectively. Line colors correspond to the number of bath metal orbitals, 
    as indicated in the top color palette. The inset in the right panel compares IESH and OSH when $n_\text{b} = 150$.}
    \label{fig:vib_hist_neq}
\end{figure}

\begin{figure}[htbp]
    \centering
    \includegraphics[width=0.45\linewidth]{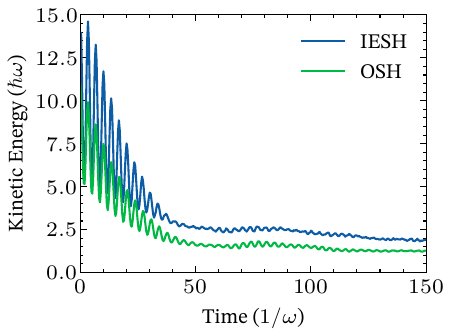}
    \caption{Kinetic energy of \ce{NO} as function of time for OSH and IESH. $n_\text{b} = 150$.}
    \label{fig:vib_hist_neq_150}
\end{figure}

Finally, we examine the convergence of \ce{NO}'s final vibrational distribution as $n_\text{b}$ increases. 
Figure~\ref{fig:vib_hist_neq} shows that the converged $\nu$ distribution peaks at $\nu=0$ and decreases 
with increasing $\nu$, a trend shared by OSH and IESH.  However, OSH predicts higher probabilities for 
lower vibrational states, indicating greater relaxation to lower kinetic energy. Specifically, OSH converges 
to a final kinetic energy of $1.25 \hbar \omega$, compared to $2 \hbar \omega$ for IESH (see Figure~\ref{fig:vib_hist_neq_150}). 
Additionally, IESH shows significantly higher kinetic energy than OSH during the transient period ($t < 50 / \omega$). 
Notably, these simulations are highly expensive, requiring $n_\text{b}>120$ and over 5000 trajectories 
to converge the $\nu$ distribution. Despite this, OSH can achieve a 1.7x times faster even when compared with a highly 
optimized IESH implementation \cite{miao18} (see Supporting Information for details).

\section{Conclusion}\label{sec:conclusion}
We have introduced Orbital Surface Hopping (OSH), a novel and efficient algorithm for modeling non-interacting multi-electron systems. 
This algorithm is motivated by the orbital Quantum-Classical Liouville Equation (o-QCLE), which is exactly derived as 
a quantum-classical approximation of the Liouville equation for the reduced many-body density matrix. OSH is applied 
to study electron and energy transfer dynamics in the molecule-metal surface, demonstrating results that are identical 
or comparable to those of the Independent Electron Surface Hopping (IESH) algorithm, but at significantly lower 
computational cost. Two scenarios are examined: for electron transfer at thermal equilibrium, OSH and IESH produced 
nearly identical results. For vibrational relaxation of highly excited molecules, OSH and IESH yielded similar 
vibrational distributions, though OSH predicted greater energy transfer from molecular vibrations to the metal 
surface. Notably, for large systems with over 100 orbitals, OSH achieves approximately 2x speedup compared to 
an efficient implementation of IESH. We believe that the improved computation efficiency of OSH would make 
it a competitive approach in real complex systems.

\begin{acknowledgement}

W. D. thanked the funding from the National Natural Science Foundation of China (no. 22361142829) and Zhejiang Provincial Natural Science Foundation (no. XHD24B0301)). Y.-T. Ma thanked project (Grant No. ZR2021MB081) supported by Shandong Provincial Natural Science Foundation and thanked Chongxiao Zhao and Yunhao Liu for a helpful second quantization theory discussion. The authors thanked the Westlake University Supercomputer Center for the facility support and technical assistance.

\end{acknowledgement}

\begin{suppinfo}

Supporting Information Available: [Potential energy surface for NO-Au(111) system 
and simulation cost comparison of OSH, IESH, FCI-SH methods] This material is
available free of charge via the Internet at http://pubs.acs.org.
\end{suppinfo}

\appendix
\section{\label{app:lvn} Derivation of Orbital Liouville-von Neumann Equation (Eq.~\ref{eqn:single_body_lvn})}

Apply $\hat{d}_i^{\dagger} \hat{d}_j$ from the right to both sides of Eq.~\ref{eqn:lvn}, and then take trace over the electronic degrees of freedom,
\begin{equation}
    \frac{\partial \hat{\sigma}_{ji}}{\partial t} = -\frac{\im}{\hbar} \left[ \Tr_\text{e}(\hat{H}_\text{total} \hat{\rho} \hat{d}_i^{\dagger} \hat{d}_j) - \Tr_\text{e}(\hat{\rho} \hat{H}_\text{total} \hat{d}_i^{\dagger} \hat{d}_j)  \right]. 
\end{equation}
Realize that $\hat{T}_\text{n}$ and $U(\hat{\bm{R}})$ commute with the creation and annihilation operators and can be moved out of the trace, hence
\begin{equation}
\begin{split}
    \frac{\partial \hat{\sigma}_{ji}}{\partial t} 
    = & -\frac{\im}{\hbar}\sum_{kl} \left[ h_{kl}(\hat{\bm{R}})\Tr_\text{e}(\hat{d}_k^{\dagger} \hat{d}_l \hat{\rho} \hat{d}_i^{\dagger} \hat{d}_j) - \Tr_\text{e}(\hat{\rho} \hat{d}_k^{\dagger} \hat{d}_l \hat{d}_i^{\dagger} \hat{d}_j) h_{kl}(\hat{\bm{R}})  \right] \\
    & -\frac{\im}{\hbar} \comm{\hat{T}_\text{n}}{\hat{\sigma}_{ij}}-\frac{\im}{\hbar} \comm{U(\hat{\bm{R}})}{\hat{\sigma}_{ij}}.
\end{split}    
\end{equation}
Then, apply the anti-commutation relation $\acomm{\hat{d}_i^{\dagger}}{\hat{d}_j} = \delta_{ij}$ and the cyclic properties of trace, we have 
\[
\begin{aligned}
    \Tr_\text{e}(\hat{d}_k^{\dagger} \hat{d}_l \hat{\rho} \hat{d}_i^{\dagger} \hat{d}_j) =
    \hat{\sigma}_{li} \delta_{jk} - \Tr_\text{e}(\hat{d}_i^{\dagger} \hat{d}_k^{\dagger} \hat{d}_j  \hat{d}_l \hat{\rho} ), \\
    \Tr_\text{e}(\hat{\rho} \hat{d}_k^{\dagger} \hat{d}_l \hat{d}_i^{\dagger} \hat{d}_j) =
    \hat{\sigma}_{jk} \delta{il}- \Tr_\text{e}(\hat{\rho} \hat{d}_k^{\dagger} \hat{d}_i^{\dagger} \hat{d}_l \hat{d}_j).
\end{aligned}
\]
The second terms of the above two equations cancel each other out since $\acomm{\hat{d}_i}{\hat{d}_j} = \acomm{\hat{d}_i^{\dagger}}{\hat{d}_j^{\dagger}} = 0$. While for the first term, swap the dummy index $k \leftrightarrow l$ for the second line leads to
\begin{align*} 
    &\sum_{kl} h_{kl}(\hat{\bm{R}})\left[ \Tr_\text{e}(\hat{d}_k^{\dagger} \hat{d}_l \hat{\rho} \hat{d}_i^{\dagger} \hat{d}_j) - \Tr_\text{e}(\hat{\rho} \hat{d}_k^{\dagger} \hat{d}_l \hat{d}_i^{\dagger} \hat{d}_j)  \right] \\
    =& \sum_{kl} \left[  h_{kl}(\hat{\bm{R}}) \hat{\sigma}_{li} \delta_{jk} - h_{lk}(\hat{\bm{R}}) \hat{\sigma}_{jl} \delta_{ik} \right], \\
    =& \sum_{l} \left[  h_{jl}(\hat{\bm{R}}) \hat{\sigma}_{li} - \hat{\sigma}_{jl} h_{li}(\hat{\bm{R}}) \right].
\end{align*}
Hence, we have proved Eq.~\ref{eqn:single_body_lvn}, i.e., the LvN for single body orbitals .

\section{\label{app:nac} Derivation of the fist order derivative coupling (DC) of different states.} 

Considered the DC between two adiabatic stats $\bm{j} \  (\ket{\bm{j}}=|j_{1}j_{2},\dots,j_{Ne}|)$ and $\bm{k}\  (\ket{\bm{k}}=|k_{1}k_{2},\dots,k_{Ne}|)$: 
\begin{equation}
  \braket{\bm{j}}{\frac{\partial\bm{k}}{\partial R}} = \frac{\bra{\bm{j}}\frac{\partial\hat{H}_{el}}{\partial R}\ket{\bm{k}}}{E_{\bm{k}}-E_{\bm{j}}}.  
\end{equation}

Use the unit projector $\bm{1} = \sum_{m}\ket{m}\bra{m}=\sum_{n}\ket{n}\bra{n}$

\begin{align*}
   &\frac{\bra{\bm{j}}\frac{\partial\hat{H}_{el}}{\partial R}\ket{\bm{k}}}{E_{\bm{k}}-E_{\bm{j}}}  \\
   =& \frac{\sum_{mn}^{N_{orb}}\braket{\bm{j}}{m}\bra{m}\frac{\partial\hat{H}_{el}}{\partial R}\ket{n}\braket{n}{\bm{k}}}{E_{\bm{k}}-E_{\bm{j}}} \\
   =& \frac{\sum_{mn}^{N_{orb}}\braket{\bm{j}}{m}d_{mn}(\epsilon_{n}-\epsilon_{m})\braket{n}{\bm{k}}}{\sum_{kj}^{N_{el}}(\epsilon_{k}-\epsilon_{j})} \\ 
   =& \frac{\sum_{mn}^{N_{orb}}\bra{\bm{j}}d_{mn}(\epsilon_{n}-\epsilon_{m})b_{m}^{+}b_{n}\ket{\bm{k}}}{\sum_{kj}^{N_{el}}(\epsilon_{k}-\epsilon_{j})} \\
   =& \frac{\sum_{mn}^{N_{orb}}d_{mn}(\epsilon_{n}-\epsilon_{m})\bra{\bm{j}}b_{m}^{+}b_{n}\ket{\bm{k}}}{\sum_{kj}^{N_{el}}(\epsilon_{k}-\epsilon_{j})} \\
   =& d_{k_{i}j_{i}}.
\end{align*}

The last step uses the orthonormality property of the adiabatic orbital. The subscript of $i$ stands for 
the $i^{th}$ electron, which is used to accomplish  the antisymmetry of electrons.

\bibliography{benchmark-iesh}

\providecommand{\latin}[1]{#1}
\makeatletter
\providecommand{\doi}
  {\begingroup\let\do\@makeother\dospecials
  \catcode`\{=1 \catcode`\}=2 \doi@aux}
\providecommand{\doi@aux}[1]{\endgroup\texttt{#1}}
\makeatother
\providecommand*\mcitethebibliography{\thebibliography}
\csname @ifundefined\endcsname{endmcitethebibliography}  {\let\endmcitethebibliography\endthebibliography}{}
\begin{mcitethebibliography}{65}
\providecommand*\natexlab[1]{#1}
\providecommand*\mciteSetBstSublistMode[1]{}
\providecommand*\mciteSetBstMaxWidthForm[2]{}
\providecommand*\mciteBstWouldAddEndPuncttrue
  {\def\EndOfBibitem{\unskip.}}
\providecommand*\mciteBstWouldAddEndPunctfalse
  {\let\EndOfBibitem\relax}
\providecommand*\mciteSetBstMidEndSepPunct[3]{}
\providecommand*\mciteSetBstSublistLabelBeginEnd[3]{}
\providecommand*\EndOfBibitem{}
\mciteSetBstSublistMode{f}
\mciteSetBstMaxWidthForm{subitem}{(\alph{mcitesubitemcount})}
\mciteSetBstSublistLabelBeginEnd
  {\mcitemaxwidthsubitemform\space}
  {\relax}
  {\relax}

\bibitem[Alec M. Wodtke~* and Auerbach(2004)Alec M. Wodtke~*, and Auerbach]{wodtke2004}
Alec M. Wodtke~*,~J. C.~T.; Auerbach,~D.~J. Electronically non-adiabatic interactions of molecules at metal surfaces: Can we trust the Born–Oppenheimer approximation for surface chemistry? \emph{International Reviews in Physical Chemistry} \textbf{2004}, \emph{23}, 513--539\relax
\mciteBstWouldAddEndPuncttrue
\mciteSetBstMidEndSepPunct{\mcitedefaultmidpunct}
{\mcitedefaultendpunct}{\mcitedefaultseppunct}\relax
\EndOfBibitem
\bibitem[White \latin{et~al.}(2005)White, Chen, Matsiev, Auerbach, and Wodtke]{Wodtke04b}
White,~J.~D.; Chen,~J.; Matsiev,~D.; Auerbach,~D.~J.; Wodtke,~A.~M. Conversion of large-amplitude vibration to electron excitation at a metal surface. \emph{Nature} \textbf{2005}, 503--505\relax
\mciteBstWouldAddEndPuncttrue
\mciteSetBstMidEndSepPunct{\mcitedefaultmidpunct}
{\mcitedefaultendpunct}{\mcitedefaultseppunct}\relax
\EndOfBibitem
\bibitem[Huang \latin{et~al.}(2000)Huang, Rettner, Auerbach, and Wodtke]{Wodtke20a}
Huang,~Y.; Rettner,~C.~T.; Auerbach,~D.~J.; Wodtke,~A.~M. Vibrational Promotion of Electron Transfer. \emph{Science} \textbf{2000}, \emph{290}, 111--114\relax
\mciteBstWouldAddEndPuncttrue
\mciteSetBstMidEndSepPunct{\mcitedefaultmidpunct}
{\mcitedefaultendpunct}{\mcitedefaultseppunct}\relax
\EndOfBibitem
\bibitem[Huang \latin{et~al.}(2000)Huang, Wodtke, Hou, Rettner, and Auerbach]{Wodtke20b}
Huang,~Y.; Wodtke,~A.~M.; Hou,~H.; Rettner,~C.~T.; Auerbach,~D.~J. Observation of Vibrational Excitation and Deexcitation for NO $(\mathit{v}\phantom{\rule{0ex}{0ex}}=\phantom{\rule{0ex}{0ex}}2)$ Scattering from Au(111): Evidence for Electron-Hole-Pair Mediated Energy Transfer. \emph{Phys. Rev. Lett.} \textbf{2000}, \emph{84}, 2985--2988\relax
\mciteBstWouldAddEndPuncttrue
\mciteSetBstMidEndSepPunct{\mcitedefaultmidpunct}
{\mcitedefaultendpunct}{\mcitedefaultseppunct}\relax
\EndOfBibitem
\bibitem[Tully(2000)]{tully2000rev}
Tully,~J.~C. Chemical Dynamics at Metal Surfaces. \emph{Annual Review of Physical Chemistry} \textbf{2000}, \emph{51}, 153--178\relax
\mciteBstWouldAddEndPuncttrue
\mciteSetBstMidEndSepPunct{\mcitedefaultmidpunct}
{\mcitedefaultendpunct}{\mcitedefaultseppunct}\relax
\EndOfBibitem
\bibitem[Bunermann \latin{et~al.}(2015)Bunermann, Jiang, Dorenkamp, Kandratsenka, Janke, Auerbach, and Wodtke]{HAdsporption-Bunermann2015}
Bunermann,~O.; Jiang,~H.; Dorenkamp,~Y.; Kandratsenka,~A.; Janke,~S.~M.; Auerbach,~D.~J.; Wodtke,~A.~M. Electron-hole pair excitation determines the mechanism of hydrogen atom adsorption. \emph{Science} \textbf{2015}, \emph{350}, 1346--1349\relax
\mciteBstWouldAddEndPuncttrue
\mciteSetBstMidEndSepPunct{\mcitedefaultmidpunct}
{\mcitedefaultendpunct}{\mcitedefaultseppunct}\relax
\EndOfBibitem
\bibitem[Kr\"{u}ger \latin{et~al.}(2016)Kr\"{u}ger, Meyer, Kandratsenka, Wodtke, and Sch\"{a}fer]{AgSurfNODiss_Krger2016}
Kr\"{u}ger,~B.~C.; Meyer,~S.; Kandratsenka,~A.; Wodtke,~A.~M.; Sch\"{a}fer,~T. Vibrational Inelasticity of Highly Vibrationally Excited {NO} on Ag(111). \emph{The Journal of Physical Chemistry Letters} \textbf{2016}, \emph{7}, 441--446\relax
\mciteBstWouldAddEndPuncttrue
\mciteSetBstMidEndSepPunct{\mcitedefaultmidpunct}
{\mcitedefaultendpunct}{\mcitedefaultseppunct}\relax
\EndOfBibitem
\bibitem[Geweke and Wodtke(2020)Geweke, and Wodtke]{AgSurfHClDiss-Geweke2020}
Geweke,~J.; Wodtke,~A.~M. Vibrationally inelastic scattering of {HCl} from Ag(111). \emph{The Journal of Chemical Physics} \textbf{2020}, \emph{153}, 164703\relax
\mciteBstWouldAddEndPuncttrue
\mciteSetBstMidEndSepPunct{\mcitedefaultmidpunct}
{\mcitedefaultendpunct}{\mcitedefaultseppunct}\relax
\EndOfBibitem
\bibitem[Lam \latin{et~al.}(2020)Lam, Soudackov, and Hammes-Schiffer]{PCET-Lam2020}
Lam,~Y.-C.; Soudackov,~A.~V.; Hammes-Schiffer,~S. Theory of Electrochemical Proton-Coupled Electron Transfer in Diabatic Vibronic Representation: Application to Proton Discharge on Metal Electrodes in Alkaline Solution. \emph{The Journal of Physical Chemistry C} \textbf{2020}, \emph{124}, 27309--27322\relax
\mciteBstWouldAddEndPuncttrue
\mciteSetBstMidEndSepPunct{\mcitedefaultmidpunct}
{\mcitedefaultendpunct}{\mcitedefaultseppunct}\relax
\EndOfBibitem
\bibitem[Lee \latin{et~al.}(2022)Lee, Jeon, Lee, and Park]{Surface-Lee2022}
Lee,~S.~W.; Jeon,~B.; Lee,~H.; Park,~J.~Y. Hot Electron Phenomena at Solid{\textendash}Liquid Interfaces. \emph{The Journal of Physical Chemistry Letters} \textbf{2022}, \emph{13}, 9435--9448\relax
\mciteBstWouldAddEndPuncttrue
\mciteSetBstMidEndSepPunct{\mcitedefaultmidpunct}
{\mcitedefaultendpunct}{\mcitedefaultseppunct}\relax
\EndOfBibitem
\bibitem[Luo \latin{et~al.}(2016)Luo, Jiang, Juaristi, Alducin, and Guo]{NiMethane-Luo2016}
Luo,~X.; Jiang,~B.; Juaristi,~J.~I.; Alducin,~M.; Guo,~H. Electron-hole pair effects in methane dissociative chemisorption on Ni(111). \emph{The Journal of Chemical Physics} \textbf{2016}, \emph{145}, 044704\relax
\mciteBstWouldAddEndPuncttrue
\mciteSetBstMidEndSepPunct{\mcitedefaultmidpunct}
{\mcitedefaultendpunct}{\mcitedefaultseppunct}\relax
\EndOfBibitem
\bibitem[Dorenkamp \latin{et~al.}(2018)Dorenkamp, Jiang, K\"{o}ckert, Hertl, Kammler, Janke, Kandratsenka, Wodtke, and B\"{u}nermann]{H-Dorenkamp2018}
Dorenkamp,~Y.; Jiang,~H.; K\"{o}ckert,~H.; Hertl,~N.; Kammler,~M.; Janke,~S.~M.; Kandratsenka,~A.; Wodtke,~A.~M.; B\"{u}nermann,~O. Hydrogen collisions with transition metal surfaces: Universal electronically nonadiabatic adsorption. \emph{The Journal of Chemical Physics} \textbf{2018}, \emph{148}, 034706\relax
\mciteBstWouldAddEndPuncttrue
\mciteSetBstMidEndSepPunct{\mcitedefaultmidpunct}
{\mcitedefaultendpunct}{\mcitedefaultseppunct}\relax
\EndOfBibitem
\bibitem[Ke \latin{et~al.}(2021)Ke, Erpenbeck, Peskin, and Thoss]{Mjunction-Ke2021}
Ke,~Y.; Erpenbeck,~A.; Peskin,~U.; Thoss,~M. Unraveling current-induced dissociation mechanisms in single-molecule junctions. \emph{The Journal of Chemical Physics} \textbf{2021}, \emph{154}, 234702\relax
\mciteBstWouldAddEndPuncttrue
\mciteSetBstMidEndSepPunct{\mcitedefaultmidpunct}
{\mcitedefaultendpunct}{\mcitedefaultseppunct}\relax
\EndOfBibitem
\bibitem[Teh \latin{et~al.}(2022)Teh, Dou, and Subotnik]{CISS-Teh2022}
Teh,~H.-H.; Dou,~W.; Subotnik,~J.~E. Spin polarization through a molecular junction based on nuclear Berry curvature effects. \emph{Physical Review B} \textbf{2022}, \emph{106}\relax
\mciteBstWouldAddEndPuncttrue
\mciteSetBstMidEndSepPunct{\mcitedefaultmidpunct}
{\mcitedefaultendpunct}{\mcitedefaultseppunct}\relax
\EndOfBibitem
\bibitem[Beck \latin{et~al.}(2000)Beck, Jäckle, Worth, and Meyer]{Beck20001}
Beck,~M.; Jäckle,~A.; Worth,~G.; Meyer,~H.-D. The multiconfiguration time-dependent Hartree (MCTDH) method: a highly efficient algorithm for propagating wavepackets. \emph{Physics Reports} \textbf{2000}, \emph{324}, 1--105\relax
\mciteBstWouldAddEndPuncttrue
\mciteSetBstMidEndSepPunct{\mcitedefaultmidpunct}
{\mcitedefaultendpunct}{\mcitedefaultseppunct}\relax
\EndOfBibitem
\bibitem[Xu \latin{et~al.}(2019)Xu, Liu, Song, and Shi]{Qiang2019}
Xu,~M.; Liu,~Y.; Song,~K.; Shi,~Q. {A non-perturbative approach to simulate heterogeneous electron transfer dynamics: Effective mode treatment of the continuum electronic states}. \emph{The Journal of Chemical Physics} \textbf{2019}, \emph{150}, 044109\relax
\mciteBstWouldAddEndPuncttrue
\mciteSetBstMidEndSepPunct{\mcitedefaultmidpunct}
{\mcitedefaultendpunct}{\mcitedefaultseppunct}\relax
\EndOfBibitem
\bibitem[Dou \latin{et~al.}(2017)Dou, Miao, and Subotnik]{joseph2017}
Dou,~W.; Miao,~G.; Subotnik,~J.~E. Born-Oppenheimer Dynamics, Electronic Friction, and the Inclusion of Electron-Electron Interactions. \emph{Phys. Rev. Lett.} \textbf{2017}, \emph{119}, 046001\relax
\mciteBstWouldAddEndPuncttrue
\mciteSetBstMidEndSepPunct{\mcitedefaultmidpunct}
{\mcitedefaultendpunct}{\mcitedefaultseppunct}\relax
\EndOfBibitem
\bibitem[Smith and Hynes(1993)Smith, and Hynes]{james1993}
Smith,~B.~B.; Hynes,~J.~T. {Electronic friction and electron transfer rates at metallic electrodes}. \emph{The Journal of Chemical Physics} \textbf{1993}, \emph{99}, 6517--6530\relax
\mciteBstWouldAddEndPuncttrue
\mciteSetBstMidEndSepPunct{\mcitedefaultmidpunct}
{\mcitedefaultendpunct}{\mcitedefaultseppunct}\relax
\EndOfBibitem
\bibitem[d'Agliano \latin{et~al.}(1975)d'Agliano, Kumar, Schaich, and Suhl]{Suhl1975}
d'Agliano,~E.~G.; Kumar,~P.; Schaich,~W.; Suhl,~H. Brownian motion model of the interactions between chemical species and metallic electrons: Bootstrap derivation and parameter evaluation. \emph{Phys. Rev. B} \textbf{1975}, \emph{11}, 2122--2143\relax
\mciteBstWouldAddEndPuncttrue
\mciteSetBstMidEndSepPunct{\mcitedefaultmidpunct}
{\mcitedefaultendpunct}{\mcitedefaultseppunct}\relax
\EndOfBibitem
\bibitem[Brandbyge \latin{et~al.}(1995)Brandbyge, Hedeg\aa{}rd, Heinz, Misewich, and Newns]{Newns1995}
Brandbyge,~M.; Hedeg\aa{}rd,~P.; Heinz,~T.~F.; Misewich,~J.~A.; Newns,~D.~M. Electronically driven adsorbate excitation mechanism in femtosecond-pulse laser desorption. \emph{Phys. Rev. B} \textbf{1995}, \emph{52}, 6042--6056\relax
\mciteBstWouldAddEndPuncttrue
\mciteSetBstMidEndSepPunct{\mcitedefaultmidpunct}
{\mcitedefaultendpunct}{\mcitedefaultseppunct}\relax
\EndOfBibitem
\bibitem[Head‐Gordon and Tully(1995)Head‐Gordon, and Tully]{HGM1995}
Head‐Gordon,~M.; Tully,~J.~C. {Molecular dynamics with electronic frictions}. \emph{The Journal of Chemical Physics} \textbf{1995}, \emph{103}, 10137--10145\relax
\mciteBstWouldAddEndPuncttrue
\mciteSetBstMidEndSepPunct{\mcitedefaultmidpunct}
{\mcitedefaultendpunct}{\mcitedefaultseppunct}\relax
\EndOfBibitem
\bibitem[Dou \latin{et~al.}(2015)Dou, Nitzan, and Subotnik]{Joseph2015}
Dou,~W.; Nitzan,~A.; Subotnik,~J.~E. {Frictional effects near a metal surface}. \emph{The Journal of Chemical Physics} \textbf{2015}, \emph{143}, 054103\relax
\mciteBstWouldAddEndPuncttrue
\mciteSetBstMidEndSepPunct{\mcitedefaultmidpunct}
{\mcitedefaultendpunct}{\mcitedefaultseppunct}\relax
\EndOfBibitem
\bibitem[Dou and Subotnik(2016)Dou, and Subotnik]{Joseph2016}
Dou,~W.; Subotnik,~J.~E. {A many-body states picture of electronic friction: The case of multiple orbitals and multiple electronic states}. \emph{The Journal of Chemical Physics} \textbf{2016}, \emph{145}, 054102\relax
\mciteBstWouldAddEndPuncttrue
\mciteSetBstMidEndSepPunct{\mcitedefaultmidpunct}
{\mcitedefaultendpunct}{\mcitedefaultseppunct}\relax
\EndOfBibitem
\bibitem[Bode \latin{et~al.}(2011)Bode, Kusminskiy, Egger, and von Oppen]{Felix2011}
Bode,~N.; Kusminskiy,~S.~V.; Egger,~R.; von Oppen,~F. Scattering Theory of Current-Induced Forces in Mesoscopic Systems. \emph{Phys. Rev. Lett.} \textbf{2011}, \emph{107}, 036804\relax
\mciteBstWouldAddEndPuncttrue
\mciteSetBstMidEndSepPunct{\mcitedefaultmidpunct}
{\mcitedefaultendpunct}{\mcitedefaultseppunct}\relax
\EndOfBibitem
\bibitem[L\"u \latin{et~al.}(2012)L\"u, Brandbyge, Hedeg\aa{}rd, Todorov, and Dundas]{Daniel2012}
L\"u,~J.-T.; Brandbyge,~M.; Hedeg\aa{}rd,~P.; Todorov,~T.~N.; Dundas,~D. Current-induced atomic dynamics, instabilities, and Raman signals: Quasiclassical Langevin equation approach. \emph{Phys. Rev. B} \textbf{2012}, \emph{85}, 245444\relax
\mciteBstWouldAddEndPuncttrue
\mciteSetBstMidEndSepPunct{\mcitedefaultmidpunct}
{\mcitedefaultendpunct}{\mcitedefaultseppunct}\relax
\EndOfBibitem
\bibitem[Persson and Persson(1980)Persson, and Persson]{PERSSON1980}
Persson,~B.; Persson,~M. Vibrational lifetime for CO adsorbed on Cu(100). \emph{Solid State Communications} \textbf{1980}, \emph{36}, 175--179\relax
\mciteBstWouldAddEndPuncttrue
\mciteSetBstMidEndSepPunct{\mcitedefaultmidpunct}
{\mcitedefaultendpunct}{\mcitedefaultseppunct}\relax
\EndOfBibitem
\bibitem[Shenvi \latin{et~al.}(2009)Shenvi, Roy, and Tully]{tully2009j}
Shenvi,~N.; Roy,~S.; Tully,~J.~C. {Nonadiabatic dynamics at metal surfaces: Independent-electron surface hopping}. \emph{The Journal of Chemical Physics} \textbf{2009}, \emph{130}, 174107\relax
\mciteBstWouldAddEndPuncttrue
\mciteSetBstMidEndSepPunct{\mcitedefaultmidpunct}
{\mcitedefaultendpunct}{\mcitedefaultseppunct}\relax
\EndOfBibitem
\bibitem[Shenvi and Tully(2012)Shenvi, and Tully]{tully2012fd}
Shenvi,~N.; Tully,~J.~C. Nonadiabatic dynamics at metal surfaces: Independent electron surface hopping with phonon and electron thermostats. \emph{Faraday Discuss.} \textbf{2012}, \emph{157}, 325--335\relax
\mciteBstWouldAddEndPuncttrue
\mciteSetBstMidEndSepPunct{\mcitedefaultmidpunct}
{\mcitedefaultendpunct}{\mcitedefaultseppunct}\relax
\EndOfBibitem
\bibitem[Malpathak and Ananth(2024)Malpathak, and Ananth]{Nandini2024}
Malpathak,~S.; Ananth,~N. A Linearized Semiclassical Dynamics Study of the Multiquantum Vibrational Relaxation of NO Scattering from a Au(111) Surface. \emph{The Journal of Physical Chemistry Letters} \textbf{2024}, \emph{15}, 794--801, PMID: 38232133\relax
\mciteBstWouldAddEndPuncttrue
\mciteSetBstMidEndSepPunct{\mcitedefaultmidpunct}
{\mcitedefaultendpunct}{\mcitedefaultseppunct}\relax
\EndOfBibitem
\bibitem[Dou \latin{et~al.}(2015)Dou, Nitzan, and Subotnik]{QCMEalg-D2015}
Dou,~W.; Nitzan,~A.; Subotnik,~J.~E. Surface hopping with a manifold of electronic states. {II}. Application to the many-body Anderson-Holstein model. \emph{The Journal of Chemical Physics} \textbf{2015}, \emph{142}, 084110\relax
\mciteBstWouldAddEndPuncttrue
\mciteSetBstMidEndSepPunct{\mcitedefaultmidpunct}
{\mcitedefaultendpunct}{\mcitedefaultseppunct}\relax
\EndOfBibitem
\bibitem[Dou \latin{et~al.}(2015)Dou, Nitzan, and Subotnik]{QCMEderive-D2015}
Dou,~W.; Nitzan,~A.; Subotnik,~J.~E. Surface hopping with a manifold of electronic states. {III}. Transients, broadening, and the Marcus picture. \emph{The Journal of Chemical Physics} \textbf{2015}, \emph{142}, 234106\relax
\mciteBstWouldAddEndPuncttrue
\mciteSetBstMidEndSepPunct{\mcitedefaultmidpunct}
{\mcitedefaultendpunct}{\mcitedefaultseppunct}\relax
\EndOfBibitem
\bibitem[Dou and Subotnik(2018)Dou, and Subotnik]{Joseph2018}
Dou,~W.; Subotnik,~J.~E. {Perspective: How to understand electronic friction}. \emph{The Journal of Chemical Physics} \textbf{2018}, \emph{148}, 230901\relax
\mciteBstWouldAddEndPuncttrue
\mciteSetBstMidEndSepPunct{\mcitedefaultmidpunct}
{\mcitedefaultendpunct}{\mcitedefaultseppunct}\relax
\EndOfBibitem
\bibitem[Wang and Jia(2021)Wang, and Jia]{Wang2021}
Wang,~Y.; Jia,~Y. A path integral approach to electronic friction of a nanometer-sized tip scanning a metal surface. \emph{Communications in Theoretical Physics} \textbf{2021}, \emph{73}, 045701\relax
\mciteBstWouldAddEndPuncttrue
\mciteSetBstMidEndSepPunct{\mcitedefaultmidpunct}
{\mcitedefaultendpunct}{\mcitedefaultseppunct}\relax
\EndOfBibitem
\bibitem[Martinazzo and Burghardt(2022)Martinazzo, and Burghardt]{Burghardt2022prl}
Martinazzo,~R.; Burghardt,~I. Quantum Dynamics with Electronic Friction. \emph{Phys. Rev. Lett.} \textbf{2022}, \emph{128}, 206002\relax
\mciteBstWouldAddEndPuncttrue
\mciteSetBstMidEndSepPunct{\mcitedefaultmidpunct}
{\mcitedefaultendpunct}{\mcitedefaultseppunct}\relax
\EndOfBibitem
\bibitem[Martinazzo and Burghardt(2022)Martinazzo, and Burghardt]{Burghardt2022pra}
Martinazzo,~R.; Burghardt,~I. Quantum theory of electronic friction. \emph{Phys. Rev. A} \textbf{2022}, \emph{105}, 052215\relax
\mciteBstWouldAddEndPuncttrue
\mciteSetBstMidEndSepPunct{\mcitedefaultmidpunct}
{\mcitedefaultendpunct}{\mcitedefaultseppunct}\relax
\EndOfBibitem
\bibitem[Spiering and Meyer(2018)Spiering, and Meyer]{Jorg2018}
Spiering,~P.; Meyer,~J. Testing Electronic Friction Models: Vibrational De-excitation in Scattering of H2 and D2 from Cu(111). \emph{The Journal of Physical Chemistry Letters} \textbf{2018}, \emph{9}, 1803--1808, PMID: 29528648\relax
\mciteBstWouldAddEndPuncttrue
\mciteSetBstMidEndSepPunct{\mcitedefaultmidpunct}
{\mcitedefaultendpunct}{\mcitedefaultseppunct}\relax
\EndOfBibitem
\bibitem[Zhang \latin{et~al.}(2020)Zhang, Maurer, and Jiang]{Bin2020jpcc}
Zhang,~Y.; Maurer,~R.~J.; Jiang,~B. Symmetry-Adapted High Dimensional Neural Network Representation of Electronic Friction Tensor of Adsorbates on Metals. \emph{The Journal of Physical Chemistry C} \textbf{2020}, \emph{124}, 186--195\relax
\mciteBstWouldAddEndPuncttrue
\mciteSetBstMidEndSepPunct{\mcitedefaultmidpunct}
{\mcitedefaultendpunct}{\mcitedefaultseppunct}\relax
\EndOfBibitem
\bibitem[Zhou \latin{et~al.}(2022)Zhou, Meng, Guo, and Jiang]{Bin2022jpcl}
Zhou,~X.; Meng,~G.; Guo,~H.; Jiang,~B. First-Principles Insights into Adiabatic and Nonadiabatic Vibrational Energy-Transfer Dynamics during Molecular Scattering from Metal Surfaces: The Importance of Surface Reactivity. \emph{The Journal of Physical Chemistry Letters} \textbf{2022}, \emph{13}, 3450--3461, PMID: 35412832\relax
\mciteBstWouldAddEndPuncttrue
\mciteSetBstMidEndSepPunct{\mcitedefaultmidpunct}
{\mcitedefaultendpunct}{\mcitedefaultseppunct}\relax
\EndOfBibitem
\bibitem[Gardner \latin{et~al.}(2023)Gardner, Habershon, and Maurer]{Reinhard2023}
Gardner,~J.; Habershon,~S.; Maurer,~R.~J. Assessing Mixed Quantum-Classical Molecular Dynamics Methods for Nonadiabatic Dynamics of Molecules on Metal Surfaces. \emph{The Journal of Physical Chemistry C} \textbf{2023}, \emph{127}, 15257--15270\relax
\mciteBstWouldAddEndPuncttrue
\mciteSetBstMidEndSepPunct{\mcitedefaultmidpunct}
{\mcitedefaultendpunct}{\mcitedefaultseppunct}\relax
\EndOfBibitem
\bibitem[Bui \latin{et~al.}(2023)Bui, Thiemann, Michaelides, and Cox]{StephenJ2023}
Bui,~A.~T.; Thiemann,~F.~L.; Michaelides,~A.; Cox,~S.~J. Classical Quantum Friction at Water–Carbon Interfaces. \emph{Nano Letters} \textbf{2023}, \emph{23}, 580--587, PMID: 36626824\relax
\mciteBstWouldAddEndPuncttrue
\mciteSetBstMidEndSepPunct{\mcitedefaultmidpunct}
{\mcitedefaultendpunct}{\mcitedefaultseppunct}\relax
\EndOfBibitem
\bibitem[Wang and Dou(2023)Wang, and Dou]{dou2023jcp}
Wang,~Y.; Dou,~W. {Nonadiabatic dynamics near metal surfaces under Floquet engineering: Floquet electronic friction vs Floquet surface hopping}. \emph{The Journal of Chemical Physics} \textbf{2023}, \emph{159}, 094103\relax
\mciteBstWouldAddEndPuncttrue
\mciteSetBstMidEndSepPunct{\mcitedefaultmidpunct}
{\mcitedefaultendpunct}{\mcitedefaultseppunct}\relax
\EndOfBibitem
\bibitem[Mosallanejad \latin{et~al.}(2023)Mosallanejad, Chen, and Dou]{dou2023prb}
Mosallanejad,~V.; Chen,~J.; Dou,~W. Floquet-driven frictional effects. \emph{Phys. Rev. B} \textbf{2023}, \emph{107}, 184314\relax
\mciteBstWouldAddEndPuncttrue
\mciteSetBstMidEndSepPunct{\mcitedefaultmidpunct}
{\mcitedefaultendpunct}{\mcitedefaultseppunct}\relax
\EndOfBibitem
\bibitem[Chen \latin{et~al.}(2024)Chen, Liu, Mosallanejad, and Dou]{dou2024jpcc}
Chen,~J.; Liu,~W.; Mosallanejad,~V.; Dou,~W. Floquet Nonadiabatic Nuclear Dynamics with Photoinduced Lorentz-like Force in Quantum Transport. \emph{The Journal of Physical Chemistry C} \textbf{2024}, \emph{128}, 11219--11228\relax
\mciteBstWouldAddEndPuncttrue
\mciteSetBstMidEndSepPunct{\mcitedefaultmidpunct}
{\mcitedefaultendpunct}{\mcitedefaultseppunct}\relax
\EndOfBibitem
\bibitem[Shenvi \latin{et~al.}(2009)Shenvi, Roy, and Tully]{tully2009s}
Shenvi,~N.; Roy,~S.; Tully,~J.~C. Dynamical Steering and Electronic Excitation in NO Scattering from a Gold Surface. \emph{Science} \textbf{2009}, \emph{326}, 829--832\relax
\mciteBstWouldAddEndPuncttrue
\mciteSetBstMidEndSepPunct{\mcitedefaultmidpunct}
{\mcitedefaultendpunct}{\mcitedefaultseppunct}\relax
\EndOfBibitem
\bibitem[Roy \latin{et~al.}(2009)Roy, Shenvi, and Tully]{tully2009jpcc}
Roy,~S.; Shenvi,~N.; Tully,~J.~C. Dynamics of Open-Shell Species at Metal Surfaces. \emph{The Journal of Physical Chemistry C} \textbf{2009}, \emph{113}, 16311--16320\relax
\mciteBstWouldAddEndPuncttrue
\mciteSetBstMidEndSepPunct{\mcitedefaultmidpunct}
{\mcitedefaultendpunct}{\mcitedefaultseppunct}\relax
\EndOfBibitem
\bibitem[Kapral and Ciccotti(1999)Kapral, and Ciccotti]{QCLE-Kapral1999}
Kapral,~R.; Ciccotti,~G. Mixed quantum-classical dynamics. \emph{The Journal of Chemical Physics} \textbf{1999}, \emph{110}, 8919–8929\relax
\mciteBstWouldAddEndPuncttrue
\mciteSetBstMidEndSepPunct{\mcitedefaultmidpunct}
{\mcitedefaultendpunct}{\mcitedefaultseppunct}\relax
\EndOfBibitem
\bibitem[Subotnik(2010)]{aug-Ehrenfest-Subotnik2010}
Subotnik,~J.~E. Augmented Ehrenfest dynamics yields a rate for surface hopping. \emph{The Journal of Chemical Physics} \textbf{2010}, \emph{132}\relax
\mciteBstWouldAddEndPuncttrue
\mciteSetBstMidEndSepPunct{\mcitedefaultmidpunct}
{\mcitedefaultendpunct}{\mcitedefaultseppunct}\relax
\EndOfBibitem
\bibitem[Subotnik \latin{et~al.}(2013)Subotnik, Ouyang, and Landry]{derive-Subotnik2013}
Subotnik,~J.~E.; Ouyang,~W.; Landry,~B.~R. Can we derive Tully’s surface-hopping algorithm from the semiclassical quantum Liouville equation? Almost, but only with decoherence. \emph{The Journal of Chemical Physics} \textbf{2013}, \emph{139}\relax
\mciteBstWouldAddEndPuncttrue
\mciteSetBstMidEndSepPunct{\mcitedefaultmidpunct}
{\mcitedefaultendpunct}{\mcitedefaultseppunct}\relax
\EndOfBibitem
\bibitem[Mannouch and Richardson(2023)Mannouch, and Richardson]{MASH-Mannouch2023}
Mannouch,~J.~R.; Richardson,~J.~O. A mapping approach to surface hopping. \emph{The Journal of Chemical Physics} \textbf{2023}, \emph{158}\relax
\mciteBstWouldAddEndPuncttrue
\mciteSetBstMidEndSepPunct{\mcitedefaultmidpunct}
{\mcitedefaultendpunct}{\mcitedefaultseppunct}\relax
\EndOfBibitem
\bibitem[Nielsen \latin{et~al.}(2000)Nielsen, Kapral, and Ciccotti]{MJSH-Nielsen2000}
Nielsen,~S.; Kapral,~R.; Ciccotti,~G. Mixed quantum-classical surface hopping dynamics. \emph{The Journal of Chemical Physics} \textbf{2000}, \emph{112}, 6543–6553\relax
\mciteBstWouldAddEndPuncttrue
\mciteSetBstMidEndSepPunct{\mcitedefaultmidpunct}
{\mcitedefaultendpunct}{\mcitedefaultseppunct}\relax
\EndOfBibitem
\bibitem[Kapral(2016)]{shperspective-Kapral2016}
Kapral,~R. Surface hopping from the perspective of quantum–classical Liouville dynamics. \emph{Chemical Physics} \textbf{2016}, \emph{481}, 77–83\relax
\mciteBstWouldAddEndPuncttrue
\mciteSetBstMidEndSepPunct{\mcitedefaultmidpunct}
{\mcitedefaultendpunct}{\mcitedefaultseppunct}\relax
\EndOfBibitem
\bibitem[Tully(1990)]{tully1990}
Tully,~J.~C. Molecular dynamics with electronic transitions. \emph{The Journal of Chemical Physics} \textbf{1990}, \emph{93}, 1061--1071\relax
\mciteBstWouldAddEndPuncttrue
\mciteSetBstMidEndSepPunct{\mcitedefaultmidpunct}
{\mcitedefaultendpunct}{\mcitedefaultseppunct}\relax
\EndOfBibitem
\bibitem[Hammes‐Schiffer and Tully(1994)Hammes‐Schiffer, and Tully]{schiffer94}
Hammes‐Schiffer,~S.; Tully,~J.~C. Proton transfer in solution: Molecular dynamics with quantum transitions. \emph{The Journal of Chemical Physics} \textbf{1994}, \emph{101}, 4657--4667\relax
\mciteBstWouldAddEndPuncttrue
\mciteSetBstMidEndSepPunct{\mcitedefaultmidpunct}
{\mcitedefaultendpunct}{\mcitedefaultseppunct}\relax
\EndOfBibitem
\bibitem[Landry \latin{et~al.}(2013)Landry, Falk, and Subotnik]{diabpop-Landry2013}
Landry,~B.~R.; Falk,~M.~J.; Subotnik,~J.~E. Communication: The correct interpretation of surface hopping trajectories: How to calculate electronic properties. \emph{The Journal of Chemical Physics} \textbf{2013}, \emph{139}\relax
\mciteBstWouldAddEndPuncttrue
\mciteSetBstMidEndSepPunct{\mcitedefaultmidpunct}
{\mcitedefaultendpunct}{\mcitedefaultseppunct}\relax
\EndOfBibitem
\bibitem[Pradhan and Jain(2022)Pradhan, and Jain]{amber2022}
Pradhan,~C.~S.; Jain,~A. Detailed Balance and Independent Electron Surface-Hopping Method: The Importance of Decoherence and Correct Calculation of Diabatic Populations. \emph{Journal of Chemical Theory and Computation} \textbf{2022}, \emph{18}, 4615--4626, PMID: 35880817\relax
\mciteBstWouldAddEndPuncttrue
\mciteSetBstMidEndSepPunct{\mcitedefaultmidpunct}
{\mcitedefaultendpunct}{\mcitedefaultseppunct}\relax
\EndOfBibitem
\bibitem[Poshusta(1991)]{Poshusta-overlap-1991}
Poshusta,~R.~D. Algebrants in many electron quantum mechanics. II. New computational algorithms. \emph{International Journal of Quantum Chemistry} \textbf{1991}, \emph{40}, 225--234\relax
\mciteBstWouldAddEndPuncttrue
\mciteSetBstMidEndSepPunct{\mcitedefaultmidpunct}
{\mcitedefaultendpunct}{\mcitedefaultseppunct}\relax
\EndOfBibitem
\bibitem[Gardner \latin{et~al.}(2023)Gardner, Corken, Janke, Habershon, and Maurer]{metal-Gardner2023}
Gardner,~J.; Corken,~D.; Janke,~S.~M.; Habershon,~S.; Maurer,~R.~J. Efficient implementation and performance analysis of the independent electron surface hopping method for dynamics at metal surfaces. \emph{The Journal of Chemical Physics} \textbf{2023}, \emph{158}\relax
\mciteBstWouldAddEndPuncttrue
\mciteSetBstMidEndSepPunct{\mcitedefaultmidpunct}
{\mcitedefaultendpunct}{\mcitedefaultseppunct}\relax
\EndOfBibitem
\bibitem[Szabó and Ostlund(1996)Szabó, and Ostlund]{szabo-1996}
Szabó,~A.; Ostlund,~N.~S. \emph{Modern quantum chemistry : introduction to advanced electronic structure theory}; Mineola (N.Y.) : Dover publications, 1996\relax
\mciteBstWouldAddEndPuncttrue
\mciteSetBstMidEndSepPunct{\mcitedefaultmidpunct}
{\mcitedefaultendpunct}{\mcitedefaultseppunct}\relax
\EndOfBibitem
\bibitem[Subotnik \latin{et~al.}(2013)Subotnik, Ouyang, and Landry]{Brian2013}
Subotnik,~J.~E.; Ouyang,~W.; Landry,~B.~R. {Can we derive Tully's surface-hopping algorithm from the semiclassical quantum Liouville equation? Almost, but only with decoherence}. \emph{The Journal of Chemical Physics} \textbf{2013}, \emph{139}, 214107\relax
\mciteBstWouldAddEndPuncttrue
\mciteSetBstMidEndSepPunct{\mcitedefaultmidpunct}
{\mcitedefaultendpunct}{\mcitedefaultseppunct}\relax
\EndOfBibitem
\bibitem[Ouyang \latin{et~al.}(2016)Ouyang, Dou, Jain, and Subotnik]{ouyang16}
Ouyang,~W.; Dou,~W.; Jain,~A.; Subotnik,~J.~E. Dynamics of Barrier Crossings for the Generalized Anderson–Holstein Model: Beyond Electronic Friction and Conventional Surface Hopping. \emph{Journal of Chemical Theory and Computation} \textbf{2016}, \emph{12}, 4178--4183, PMID: 27564005\relax
\mciteBstWouldAddEndPuncttrue
\mciteSetBstMidEndSepPunct{\mcitedefaultmidpunct}
{\mcitedefaultendpunct}{\mcitedefaultseppunct}\relax
\EndOfBibitem
\bibitem[Miao \latin{et~al.}(2018)Miao, Ouyang, and Subotnik]{miao18}
Miao,~G.; Ouyang,~W.; Subotnik,~J. A comparison of surface hopping approaches for capturing metal-molecule electron transfer: A broadened classical master equation versus independent electron surface hopping. \emph{The Journal of Chemical Physics} \textbf{2018}, \emph{150}, 041711\relax
\mciteBstWouldAddEndPuncttrue
\mciteSetBstMidEndSepPunct{\mcitedefaultmidpunct}
{\mcitedefaultendpunct}{\mcitedefaultseppunct}\relax
\EndOfBibitem
\bibitem[Sun and Hase(2010)Sun, and Hase]{wigner_function_Sun2010}
Sun,~L.; Hase,~W.~L. Comparisons of classical and Wigner sampling of transition state energy levels for quasiclassical trajectory chemical dynamics simulations. \emph{The Journal of Chemical Physics} \textbf{2010}, \emph{133}\relax
\mciteBstWouldAddEndPuncttrue
\mciteSetBstMidEndSepPunct{\mcitedefaultmidpunct}
{\mcitedefaultendpunct}{\mcitedefaultseppunct}\relax
\EndOfBibitem
\bibitem[Meng and Jiang(2022)Meng, and Jiang]{pes-Meng2022}
Meng,~G.; Jiang,~B. A pragmatic protocol for determining charge transfer states of molecules at metal surfaces by constrained density functional theory. \emph{The Journal of Chemical Physics} \textbf{2022}, \emph{157}\relax
\mciteBstWouldAddEndPuncttrue
\mciteSetBstMidEndSepPunct{\mcitedefaultmidpunct}
{\mcitedefaultendpunct}{\mcitedefaultseppunct}\relax
\EndOfBibitem
\bibitem[Meng \latin{et~al.}(2024)Meng, Gardner, Hertl, Dou, Maurer, and Jiang]{bin24}
Meng,~G.; Gardner,~J.; Hertl,~N.; Dou,~W.; Maurer,~R.~J.; Jiang,~B. First-Principles Nonadiabatic Dynamics of Molecules at Metal Surfaces with Vibrationally Coupled Electron Transfer. \emph{Phys. Rev. Lett.} \textbf{2024}, \emph{133}, 036203\relax
\mciteBstWouldAddEndPuncttrue
\mciteSetBstMidEndSepPunct{\mcitedefaultmidpunct}
{\mcitedefaultendpunct}{\mcitedefaultseppunct}\relax
\EndOfBibitem
\end{mcitethebibliography}

\end{document}